\documentclass[conference]{IEEEtran}

\makeatletter
\def\ps@headings{%
\def\@oddhead{\mbox{}\scriptsize\rightmark \hfil \thepage}%
\def\@evenhead{\scriptsize\thepage \hfil \leftmark\mbox{}}%
\def\@oddfoot{}%
\def\@evenfoot{}}
\makeatother
\IEEEoverridecommandlockouts
\usepackage{amsmath,amssymb,amsfonts}
\usepackage{graphicx,textcomp}
\usepackage{xcolor,colortbl}
\def\BibTeX{{\rm B\kern-.05em{\sc i\kern-.025em b}\kern-.08em
    T\kern-.1667em\lower.7ex\hbox{E}\kern-.125emX}}

\usepackage[T1]{fontenc}
\usepackage[utf8]{inputenc}
\usepackage{csquotes,enumitem,tabularx,subcaption,uulm-logos,tabu,tabto}
\usepackage{algorithm,algpseudocode,xspace}
\usepackage[pdfkeywords={network reconnaissance; protocol reverse engineering; vulnerability research;}]{hyperref}
\usepackage{booktabs}

\usepackage[textsize=footnotesize,disable]{todonotes} 

\usepackage{tikz,amsmath,multirow}

\usetikzlibrary{positioning, fit, calc, backgrounds, shapes}

\usepackage[backend=biber,sortcites=true,minnames=5,maxnames=10,mincitenames=1,maxcitenames=2,bibencoding=utf8,
doi=false,isbn=false,natbib]{biblatex}
\AtEveryBibitem{%
	\clearlist{language}%
    \clearfield{doi}%
    \clearfield{pages}%
	\ifentrytype{article}{%
		\clearfield{isbn}%
	}{}%
	\ifentrytype{techreport}{%
	}{%
		\clearfield{urlyear}%
	} %
	\ifentrytype{inproceedings}{%
		\clearfield{pages}%
		\clearlist{location}%
		\clearfield{month}%
        \clearlist{publisher}%
        \iffieldundef{series}{}{\clearfield{booktitle}}%
	}{%
	}%
	\ifentrytype{misc}{%
	}{%
		\ifentrytype{online}{%
		}{%
			\clearfield{url}%
			\clearfield{urldate}%
		}
	}
}
\bibliography{PRE}

\setlist{noitemsep}

\def\BibTeX{{\rm B\kern-.05em{\sc i\kern-.025em b}\kern-.08emT\kern-.1667em\lower.7ex\hbox{E}\kern-.125emX}}

\newcolumntype{t}{>{\tt}r}

\newcommand{\tyl}{\textsc{Neme}TYL\xspace}
\newcommand{\tylrefine}{baseline\xspace}
\newcommand{\sys}{\textsc{Neme}SYS\xspace}
\newcommand{\pca}{\textsc{Neme}PCA\xspace}

\newcommand{\Land}{\ensuremath{\hspace{3ex}\land\hspace{3ex}}}

\newcommand{\datamatrix}{\ensuremath{\mathbf{X}}}
\newcommand{\covmatrix}{\ensuremath{\mathbf{C}}}

\newcommand{\loading}{\ensuremath{\mathbf{w}}}
\newcommand{\loadingi}{\ensuremath{\loading_i}}  
\newcommand{\score}{\ensuremath{\lambda}}  

\newcommand{\screeThresh}{\ensuremath{q_s}}

\newcommand{\screeMinThresh}{\ensuremath{p_s}}
\newcommand{\maxAbsolutePrincipals}{\ensuremath{p_p}}
\newcommand{\principalCountThresh}{\ensuremath{p_q}}
\newcommand{\maxLengthDeltaRatio}{\ensuremath{p_l}}
\newcommand{\minSegLen}{\ensuremath{p_c}} 
\newcommand{\contributionRelevant}{\ensuremath{p_r}}  
\newcommand{\pcDeltaMin}{\ensuremath{p_d}}  
\newcommand{\maxContribution}{\ensuremath{m}}  
\newcommand{\relaxedNZ}{p_z}  
\newcommand{\relaxedNZlength}{p_b}
\newcommand{\relaxedNoteableContrib}{p_t}

\newcommand{\kneedle}{\ensuremath{K(\lambda)}}

\begin{document}

\title{Refining Network Message Segmentation with Principal Component Analysis%
}

\author{%
\IEEEauthorblockN{Stephan Kleber}
\IEEEauthorblockA{\textit{Institute of Distributed Systems} \\
\textit{Ulm University}\\
Ulm, Germany \\
ORCID 0000-0001-9836-4897}
\and
\IEEEauthorblockN{Frank Kargl}
\IEEEauthorblockA{\textit{Institute of Distributed Systems} \\
\textit{Ulm University}\\
Ulm, Germany \\
ORCID 0000-0003-3800-8369}



%
}

\maketitle


\IEEEpubid{\begin{minipage}{\textwidth}\ \\[24pt]
        \copyright 2020 IEEE. Personal use of this material is permitted. Permission from IEEE must be
        obtained for all other uses, in any current or future media, including
        reprinting/republishing this material for advertising or promotional purposes, creating new
        collective works, for resale or redistribution to servers or lists, or reuse of any copyrighted
        component of this work in other works.\\
        IEEE Conference on Communications and Network Security 2022.
\end{minipage}} 

\IEEEpubidadjcol
\begin{abstract}
Reverse engineering of undocumented protocols is a common task in security analyses of networked services.
The communication itself, captured in traffic traces, contains much of the necessary information to perform such a protocol reverse engineering.
The comprehension of the format of unknown messages is of particular interest for binary protocols that are not human-readable.
One major challenge is to discover probable fields in a message as the basis for further analyses.
Given a set of messages, split into segments of bytes by an existing segmenter, we propose a method to refine the approximation of the field inference.
We use principle component analysis (PCA) to discover linearly correlated variance between sets of message segments.
We relocate the boundaries of the initial coarse segmentation to more accurately match with the true fields.
We perform different evaluations of our method to show its benefit for the message format inference and subsequent analysis tasks from literature that depend on the message format.
We can achieve a median improvement of the message format accuracy across different real-world protocols by up to 100\,\%.
\end{abstract}


\pdfinfo{info}

\section{Introduction}

Analyzing the threat posed by botnet and malware communication~\cite{cho_inference_2010}, validating the correct and secure design and implementation of network services~\cite{wen_protocol_2017}, and efficiently configuring smart fuzzers~\cite{gascon_pulsar:_2015}
requires the understanding of the involved network protocols.
In case of malware and proprietary systems, the protocols are often undocumented and first require protocol reverse engineering (PRE) to uncover data exfiltration or vulnerabilities in the network services.
As an example, PRE recently played an important role in discovering a severe vulnerability in the proprietary Apple Wireless Direct Link (AWDL) protocol stack~\cite{stute_one_2018}.
The vulnerability led to a zero-click exploit~\cite{beer_google_2020} affecting all of Apple's iOS-based product lines and could be fixed due to the protocol analysis.

Samples of unknown protocols can typically be collected from observing communication of devices implementing this protocol.
PRE can make use of these traffic traces to infer the specification of the unknown network protocol and thus to gain knowledge about its syntax and behavior.
The approximation of protocol fields for an unknown message syntax is required to determine the message format,
semantics of the fields' data representations, and message types.
%
Among others, protocols used for embedded systems often are optimized for efficiency and thus transmit binary data instead of ASCII-encoded, human-readable values.
The latter are called textual protocols.
Most existing PRE approaches only support textual protocols using techniques from natural language processing (NLP)~\cite{wang_semantics_2012,luo_position-based_2013}.
This requires repeated keywords and separators to search for in the structure of the messages.
The analysis of binary protocols that do not exhibit such structural features is considerably harder than of textual protocols~\cite{cui_discoverer:_2007, bermudez_towards_2016, kleber_survey_2019, duchene_state_2018}.

Widespread early approaches designed for binary protocols~\cite{kleber_survey_2019} use sequence alignment~%
\cite{bossert_towards_2014, cui_discoverer:_2007}.
Designed for bio-informatics, it solves the problem of inferring structure from a small number of sequences of amino acids.
The challenge is reduced by additional knowledge about the chemical properties of the sequences to be aligned.
For binary protocols, a large number of messages, i.\,e., sequences, of one protocol is beneficial for observing the variability of values.
The lack of conclusive properties to guide the alignment process and the larger number of sequences to be compared pose a major obstacle for applying sequence alignment to network protocols.
Therefore, more recent approaches rely on statistical variance analysis which offers improved performance.
%
We previously proposed \sys~\cite{kleber_nemesys:_2018}, one such statistical variance analysis for the message format inference of binary protocols, and
a set of refinement methods to improve the accuracy of the approximation of message fields as part of \tyl~\cite{kleber_message_2020}.

\IEEEpubidadjcol
Based on these previous works, the \textbf{main contributions} of this paper are two novel methods for segmenting and refining message formats of binary protocols:
We propose a 
\textbf{segmentation refinement} and also derive a 
\textbf{new segmenter} that works independent of \sys.
While previous work typically requires to select protocol-dependent analysis parameters, which is difficult for an unknown protocol, our refinement and segmenter are independent of any parameters.
We use the message format quality measure FMS~\cite{kleber_nemesys:_2018} to compare the existing
refinements~\cite{kleber_message_2020} to the methods of this paper.
The evaluation results show that, in most cases, we can improve the quality of the segmentation alone by about 50\,\% across different protocols, and we can significantly improve analyses relying on this segmentation, like message type identification and semantic analysis.
From the evaluation results, we deduce which of the presented methods is suited best for different tasks of PRE.


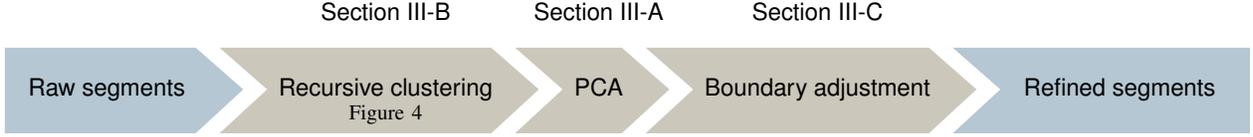
\begin{figure*}
	\centering\small
	\begin{tikzpicture}[
	every node/.style={
		outer sep=.5ex, inner sep=2.2ex,
		font=\sffamily, align=left, anchor=base}  
	]
	\coordinate (top) at (0, 2em);
	\coordinate (bot) at (0,-1.6em);
	
	\node[] at (0,0)                                                                 (raw)        {Raw segments};
	\node[right=1.5em of raw,      label={above:\autoref{sec:recursive-clustering}}] (clustering) {Recursive clustering};
	\node[right=1em of clustering, label={above:\autoref{sec:PCA}}]                  (pca)        {PCA\vphantom{g}};
	\node[right=1em of pca,        label={above:\autoref{sec:interpretation-of-the-pca}}]  (actions)    {Boundary adjustment};  
	\node[right=1.5em of actions]                                                    (refined)    {Refined segments};
    
    \node[below=-5ex of clustering, font=\footnotesize] {\autoref{fig:preparation}};
	
	\begin{scope}[on background layer, 
	every path/.style={transform canvas={xshift={.5ex}}, fill=uulm-akzent!60}
	]
	\foreach \x in {raw} {
		\path[fill=uulm!60]
		(\x.west |- top) to (\x.west |- bot)
		to ([xshift=-1em]\x.east |- bot) to ([xshift=+1em]\x.east)
		to ([xshift=-1em]\x.east |- top) to cycle;
	}
	\foreach \x in {clustering,pca,actions} {
		\path
		([xshift=-1.5em]\x.west |- top) to ([xshift=+.5em]\x.west) to
		([xshift=-1.5em]\x.west |- bot)
		to ([xshift=-1.5em]\x.east |- bot) to ([xshift=+.5em]\x.east)
		to ([xshift=-1.5em]\x.east |- top) to cycle;
	}
	\foreach \x in {refined} {
		\path[fill=uulm!60]
		([xshift=-2em]\x.west |- top) to (\x.west)
		to ([xshift=-2em]\x.west |- bot)
		to (\x.east |- bot)
		to (\x.east |- top) to cycle;
	}
	\end{scope}
	
	\end{tikzpicture}
	\caption{PCA process overview}
	\label{fig:process-overview}
\end{figure*}

\section{Related Work}
\label{sec:related}

Several surveys~\cite{narayan_survey_2015, duchene_state_2018, sija_survey_2018, kleber_survey_2019} discuss the plethora of approaches of PRE and also have proposed to structure the overall PRE process into different phases~\cite{duchene_state_2018, kleber_survey_2019}.
These phases are data collection into traces, feature extraction, message format inference, message type identification, semantic deduction, and behavior model reconstruction.
As previous work showed, support for performing the tasks of the phases on binary protocols is severely limited in comparison to textual protocols~\cite{cui_discoverer:_2007, bermudez_towards_2016, duchene_state_2018, kleber_survey_2019}.
While most phases are well covered in literature, approaches for segmenting messages have gained only limited attention despite its relevance for all subsequent phases.
We adopt the term \enquote{segment} from previous work.
A segment is a subsequence of bytes of a message that approximates a field's boundaries from the protocol definition in terms of byte positions.
In an optimal inference, the segments match the true field boundaries from the unknown syntax.

Netzob~\cite{bossert_towards_2014}, Discoverer~\cite{cui_discoverer:_2007}, and others~\cite{leita_scriptgen:_2005} deduce fields as a by-product of sequence alignment with the already mentioned disadvantages.
Existing statistical methods
either require an already existing segmentation~\cite{bossert_towards_2014,cui_discoverer:_2007,leita_scriptgen:_2005} or expect field boundaries at globally fixed positions~\cite{sun_unsupervised_2019, bermudez_towards_2016, trifilo_traffic_2009}, limiting the applicability to protocols specifically designed without variable length fields.
If meta-data and common offsets of values in messages are available, the task is as simple as finding the corresponding or correlating values in the messages.
For instance, analyzing wireless communication of a medical device or sensor node, there is no encapsulation present from which to extract addresses and identifiers.
Thus, we argue that no convincing method exists for accurately inferring structure of binary protocols~\cite{kleber_survey_2019}.


\medskip
Opposed to all previous approaches, we make very few assumptions, thus, creating a more generic approach.
It does not require a specific message format or protocol structure, like globally common byte offsets or typical field lengths.
It works without any preceding classification of messages or the identification of flows, and thus also works for protocols without any encapsulation like TCP, UDP, and IP%
\footnote{Transmission Control Protocol (RFC 793), User Datagram Protocol (RFC 768), and Internet Protocol (RFC 791)}.
The last property is of particular interest when reverse engineering proprietary wireless point-to-point protocols were no encapsulation is present or accessible, like such used by mobile devices~\cite{stute_one_2018} and for medical devices~\cite{marin_feasibility_2016, rios_understanding_2018}, simple Internet-of-Things nodes \cite{pohl_universal_2018}, or vehicle sensors, like tire-pressure monitoring systems~\cite{rouf_security_2010}.

%

In this paper, we propose refinements that enhance \sys~\cite{kleber_nemesys:_2018} 
and we compare our novel segmenter to the refinements from \tyl~\cite{kleber_message_2020}.
Our approach uses a concept for the clustering of segments that we proposed earlier~\cite{kleber_network_2022}.
Thus, we briefly introduce these three fundamental approaches in the rest of this section.


We utilize the \sys segmentation~\cite{kleber_nemesys:_2018} for a first, raw approximation of the true message structure without knowledge of the specification.
The method derives segments from the distribution of value changes within subsequent bytes of single messages.
Unlike sequence alignment or statistical methods, \sys does not compare different messages with each other.
%
It is known that \sys boundaries regularly exhibit an off-by-one error~\cite{kleber_nemesys:_2018}.
We confirmed for all of our test protocols that the vast majority of \sys' \enquote{near-match} boundary errors are such off-by-one errors.
We tried to find patterns in these errors that would enable us to create a systematic correction but could not find constant off-by-one error rules for \sys segments.
Therefore, we propose our refinement to correct this type of errors dynamically.


The approach \tyl~\cite{kleber_message_2020} identifies message types from clustering of messages by their similarity.
It additionally contains refinements to apply to the segments obtained from \sys to more correctly approximate fields.
%
%
\tyl proposes a simple frequency analysis to determine the most common segment values, which point out probable field boundaries throughout the protocol under analysis.
Furthermore, \tyl recognizes char sequences embedded within the binary protocol.
%
%


To obtain sets of comparable, related segments, which we subsequently intend to analyze together, we cluster segments based on their dissimilarity.
We base our clustering approach on our previous proposal~\cite{kleber_network_2022} to use 
DBSCAN~\cite{ester_density-based_1996}.
The used measure is the Canberra dissimilarity~\cite{kleber_message_2020} which extends the better-known Canberra distance~\cite{lance_computer_1966} to vectors of differing dimensions.
We use the Canberra dissimilarity between segments as affinity measure to guide the clustering.
%
The clustering results in concise sets of segments that overlay best, matching each others' values measured by the smallest possible Canberra dissimilarity of segments.

\addtolength{\topmargin}{.03in}
\section{Byte-wise Segment Variance Analysis}
\label{sec:BSVA}


The evaluation of different aspects of \sys~\cite{kleber_nemesys:_2018,kleber_network_2022,kleber_message_2020} has proven that it can yield useful approximations of the protocol message structure of unknown protocols.
It is highly efficient and refinements could improve some shortcomings of \sys' raw segmentation.
However, none of these proposed enhancements take variance of values within and between messages into account.

\textbf{Inter- and intra-message variance} reveals details about the message structure that is not visible otherwise and has the potential to increase the accuracy of inferred field boundaries.
We determine which bytes vary together in sets of segments that we interpret as data vectors to identify probable field boundaries.
We separate vector components, variance-locked to each other, from linearly independent vector components by Principal Component Analysis (PCA).
PCA shows which bytes from each set of segments are associated with each other and, thus, belong to a common field.

We first describe the core of our variance analysis based on PCA, in \autoref{sec:PCA}.
As illustrated in \autoref{fig:process-overview}, before we can start the PCA, we need to prepare groups of segments by a \textbf{recursive clustering} step (\autoref{sec:recursive-clustering}) and after the PCA, we interpret its results (\autoref{sec:interpretation-of-the-pca}) to \textbf{adjust the boundaries} 
of the raw input segments.
%
The whole method, which we call \emph{PCA refinement}, is embedded in a chain of other refinements, partly from related work and partly by methods we introduce in this paper.
We locate all refinements in the context of the processing chain in \autoref{sec:implementation}.

\subsection{Principal Component Analysis}
\label{sec:PCA}


The PCA calculates the multivariate main direction of variance and its scale.  
The multivariate variance identifies the components that vary together, i.\,e., linearly dependent.
Typically, PCA is applied to feature vectors and its result is used to classify samples, represented by these vectors, based on commonalities in the variance revealed by the PCA.
In contrast, we use the immediate result of a PCA to determine the contribution of different components to the variance of the vectors and thus the variance contained in a set of segments represented by the vectors.

%
%
\begin{figure}[b]
	\begin{equation}
	\datamatrix = \left(
    \footnotesize
	\begin{array}{ttttt}
	00 & 08 & 50 & 00 & 02 \\
	01 & 08 & 90 & 00 & 04 \\
	01 & 08 & 90 & 00 & 07 \\
	01 & 08 & b0 & 00 & 02 \\
	02 & 90 & 40 & 01 & 02 \\
	02 & 90 & 40 & 01 & 02 \\
	01 & 08 & 80 & 00 & 04 \\
	01 & 08 & 80 & 00 & 04
	\end{array}
	\right)
	\end{equation}
	
	\begin{equation}
	\covmatrix = \left(
	\footnotesize
	\begin{array}{rrrrr}
	0.41 &    34      &    -9.71 &   0.25     &  -0.19 \\
	34        &  3963   & -2020    &  29.14   & -53.42   \\
	-9.71  & -2020   &  1737    & -14.85   &  34.85   \\
	0.25     &    29.14 &   -14.85  &   0.21 &  -0.39 \\
	-0.19 &   -53.42 &    34.85  &  -0.39 &   3.12    \\
	\end{array}
	\right)
	\end{equation}
	\caption{Example for $\datamatrix$, with each row representing one segment's sequence of byte values, and the $\covmatrix$ of this $\datamatrix$.} 
	\label{fig:datamatrix-covmatrix}
\end{figure}
%
%
%
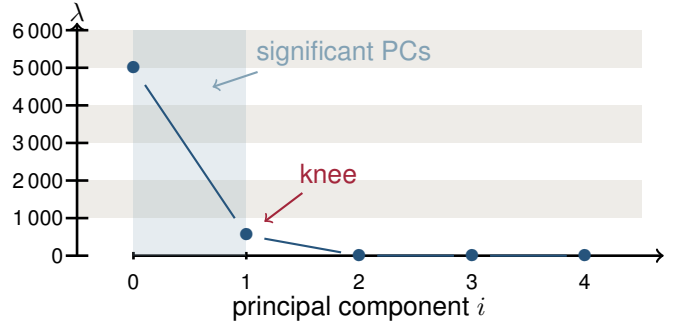
\begin{figure}
	\newcommand{\datadot}{\color{uulm-med}\large$\bullet$}

	\begin{tikzpicture}[
	yscale=0.5, xscale=1.5,
	every path/.style={line width = 1pt},
	every node/.style={font={\sffamily}}
	]
	%
	\foreach \x in {1, 3, 5}
	\draw[fill=uulm-akzent!20, draw=none] (-0.5,\x) rectangle +(5,1);
	
	\draw[->] (0.0,0.0) to (4.7,0.0);
	\draw[->] (-0.5,0.0) to (-0.5,6.2);
	\node[yshift=-2em] at (2,0) {principal component $i$};
	\node at (-0.5,6.5) {$\lambda$};
	
	\foreach \x in {0,1,2,3,4}
	\draw (\x,-0.1) -- (\x,0.1) node [below=1ex] {\footnotesize \x};
	
	\foreach \x/\y in {0/0, 1/1\,000, 2/2\,000, 3/3\,000, 4/4\,000, 5/5\,000, 6/6\,000}
	\draw (-0.6,\x) -- (-0.4,\x) node [left, inner sep=2ex] {\footnotesize \y};

	\fill[uulm, opacity=.2] (0,0) rectangle (1,6);
	\node[uulm, anchor=north west] (signpc) at (1,6) {significant PCs};
	\draw[->, thick, uulm] ($(signpc.south west)+(0.1,0.1)$) to +(-0.4,-0.4);

	\node at (0,5)     (dd0) {\datadot};
	\node at (1,0.54)  (dd1) {\datadot};
	\node at (2,0.002) (dd2) {\datadot};
	\node at (3,0)     (dd3) {\datadot};
	\node at (4,0)     (dd4) {\datadot};
	
	\draw[uulm-med, thick] (dd0) to (dd1);
	\draw[uulm-med, thick] (dd1) to (dd2);
	\draw[uulm-med, thick] (dd2) to (dd3);
	\draw[uulm-med, thick] (dd3) to (dd4);

	\node[uulm-in, above right=3ex of dd1] (knee) {knee};
	
	\draw[->, uulm-in, thick] (knee) to (dd1);
	\end{tikzpicture}
	\caption{Scree graph of PCs sorted by their eigenvalues/component scores $\score_i$.}
	\label{fig:scree}
\end{figure}
We first revisit the elements of the well-known generic PCA to clarify our usage of terms and notations.
%
%
The data vectors are collected into a matrix $\datamatrix$ with each vector as row, in our case corresponding to the bytes of one segment. 
The variance of $\datamatrix$ is described by the covariance matrix $\covmatrix$.
The equations in \autoref{fig:datamatrix-covmatrix} give an example for matrices $\datamatrix$ and $\covmatrix$ derived from segments. 
The eigenvalues (\enquote{factors} or \enquote{component scores} $\score$) and eigenvectors (\enquote{loadings} $\loadingi$) of the covariance matrix $\covmatrix$ are the foundation of the PCA.
The \emph{first} principal component (PC) is the highest eigenvalue $\lambda$ and its associated eigenvector $\loadingi$ and it intuitively states the direction of the prevalent variance in the data.
\emph{Further} PC loadings are orthogonal to the previous ones.
We call the variance at a specific byte position the \emph{notable contribution} if the $i$th eigenvector component $\loadingi$ is significantly different from zero.
According to common analysis methods, the transition between the PCs with a significant contribution and negligible components is marked by the knee of the scree graph of the eigenvalues $\lambda$ as illustrated in \autoref{fig:scree}.
%
We determine the knee by the Kneedle algorithm~\cite{satopaa_finding_2011}.

\medskip
To be able to start a PCA, two basic prerequisites must be met.
First, we need to prepare a set of segments that contain related data, so that the PCA does not measure arbitrary variance, but only meaningfully comparable variances of segments that represent the same kind of data.
Second, we need to determine which PCs significantly contribute to the variance of the data, i.\,e., the segments represented in $\datamatrix$.
\medskip

\subsubsection{Overlaying Segment Vectors and Calculating PCs}
\label{sec:align}

PCA requires an existing coarse approximation of fields, i.\,e., initial segments, to overlay the segments that are related in one set.
This overlaying is necessary to calculate the covariance from it.
We use the Canberra dissimilarity~\cite{kleber_message_2020} to find the best fitting overlay of multiple segments.
%
Thus, we superimpose the segments at the most probable useful offsets, resulting in overlays of the most meaningfully comparable message parts.
While similar to the longest common subsequence~\cite{smith_identification_1981}, we additionally allow for variations instead of requiring identical subsequences.
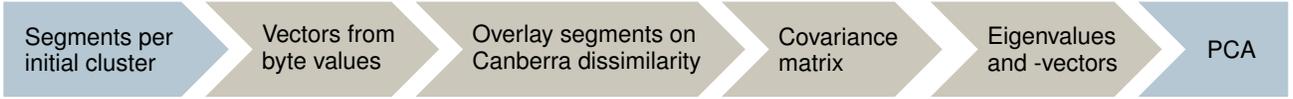
\begin{figure*}
	\centering\small
	\hspace{-2ex}
	\begin{tikzpicture}[
	every node/.style={
		outer sep=.5ex, inner sep=2ex,
		font=\sffamily, align=left}  
	]
	\coordinate (top) at (0, 2em);
	\coordinate (bot) at (0,-2em);
	
	\node[] at (0,0)               (raw) {Segments per\\ initial cluster};
	\node[right=1.5em of raw]      (vectors) {Vectors from\\ byte values};  
	\node[right=1em of vectors]    (align) {Overlay segments on\\ Canberra dissimilarity};
	\node[right=1em of align]      (covariance) {Covariance\\ matrix};
	\node[right=1.5em of covariance]  (ev) {Eigenvalues\\ and -vectors};
	\node[right=1.5em of ev]       (pca) {PCA};
	
	\begin{scope}[on background layer, 
	every path/.style={transform canvas={xshift={.5ex}}, fill=uulm-akzent!60}
	]
	\foreach \x in {raw} {
		\path[fill=uulm!60]
		(\x.west |- top) to (\x.west |- bot)
		to ([xshift=-1em]\x.east |- bot) to ([xshift=+1em]\x.east)
		to ([xshift=-1em]\x.east |- top) to cycle;
	}
	\foreach \x in {vectors,align,covariance,ev} {
		\path
		([xshift=-1.5em]\x.west |- top) to ([xshift=+.5em]\x.west) to
		([xshift=-1.5em]\x.west |- bot)
		to ([xshift=-1.5em]\x.east |- bot) to ([xshift=+.5em]\x.east)
		to ([xshift=-1.5em]\x.east |- top) to cycle;
	}
	\foreach \x in {pca} {
		\path[fill=uulm!60]
		([xshift=-2em]\x.west |- top) to (\x.west)
		to ([xshift=-2em]\x.west |- bot)
		to (\x.east |- bot)
		to (\x.east |- top) to cycle;
	}
	\end{scope}
	
	\end{tikzpicture}
	\caption{PCA preparation process performed per initial cluster.}
	\label{fig:preparation}
\end{figure*}
As \autoref{fig:preparation} illustrates, we next calculate the covariance matrix $\covmatrix$ from the aligned segment data $\datamatrix$.
Then we calculate the eigenvalues $\score$ and eigenvectors $\loadingi$ of $\covmatrix$, which the PCA uses directly.

%

\medskip
\subsubsection{Process Overview}

The PCA is \textbf{embedded} into a process that \textbf{prepares} the segment data and \textbf{interprets} the PCA result.
Introduced in \autoref{fig:process-overview} in high-level abstraction, the process is composed of the following steps in detail:
\begin{enumerate}[noitemsep]
	\item Cluster the raw segments by similarity,
	\item interpret the segment byte values as vectors,
		overlay the vectors \emph{per cluster}, 
		calculate the covariance, and from it the 
		eigenvalues and eigenvectors that are required for the subsequent PCA (summarized in \autoref{fig:preparation}),
	\item check if prerequisites for PCA are met by each cluster, 
	\item for clusters that fail this check, recurse from step 1,
	\item for clusters that pass this check, \textbf{perform the PCA}, and
	\item finally apply rules for which variance characteristics quantified by the PCA indicate field boundaries.
\end{enumerate}
The rest of this section explains the auxiliary algorithms for the preparation and interpretation of the PCA.
These algorithms make use of thresholds and other parameters for which we empirically determined suitable values (\autoref{tab:parameters}).

\subsection{Recursive Clustering}\label{sec:recursive-clustering}

Before performing the PCA, we obtain sets of comparable, related segments by clustering the segments~\cite{kleber_network_2022}. 
To adjust the clusters optimally for applying PCA, we recursively cluster segments and check whether the prerequisites for a component analysis are met by each cluster.
If a cluster does not allow to conduct a PCA, we sub-cluster it and test whether the smaller clusters result in more suitable sets of segments.
If PCA is still not applicable to a cluster, we recurse the sub-clustering on the respective cluster, otherwise we stop the recursion for this sub-cluster branch and perform the PCA on it (see \autoref{sec:PCA}).

We estimate whether a cluster is adequate for PCA by \textbf{PCA Prerequisites}. 
The main criterion is that the variance is systematic.
%
%
    We distinguish systematic from random variance by the number of significant PCs.
	For a successful PCA only a limited number of significant PCs are allowed, 
    since only if the variance is concentrated in a small number of PCs we can deduce that the data is non-random.
	We define an absolute $\maxAbsolutePrincipals$ and a relative maximum $\principalCountThresh$ of allowed significant PCs.
	The number of allowed significant PCs is exceeded if
	$$\left|\langle \score_i: \score_i > \screeThresh \rangle_{i=0}^n \right| > \min(\maxAbsolutePrincipals, \dim \score \cdot \principalCountThresh)$$
	with the threshold $\screeThresh$ of any eigenvalue $\score_i$ for a PC to count as significant:
	$\screeThresh = \min( \kneedle, \frac{1}{10} \score_0, \screeMinThresh )$.
If the condition is true and thus the PCA will fail on the cluster, we further recurse the sub-clustering to gain a subset of segments that then is appropriate.
%
%
We now interpret the PCA of each cluster.

\begin{figure}[!b]
	\includegraphics[width=\linewidth, trim=2cm 0.8cm 1.6cm 1.4cm, clip]
		{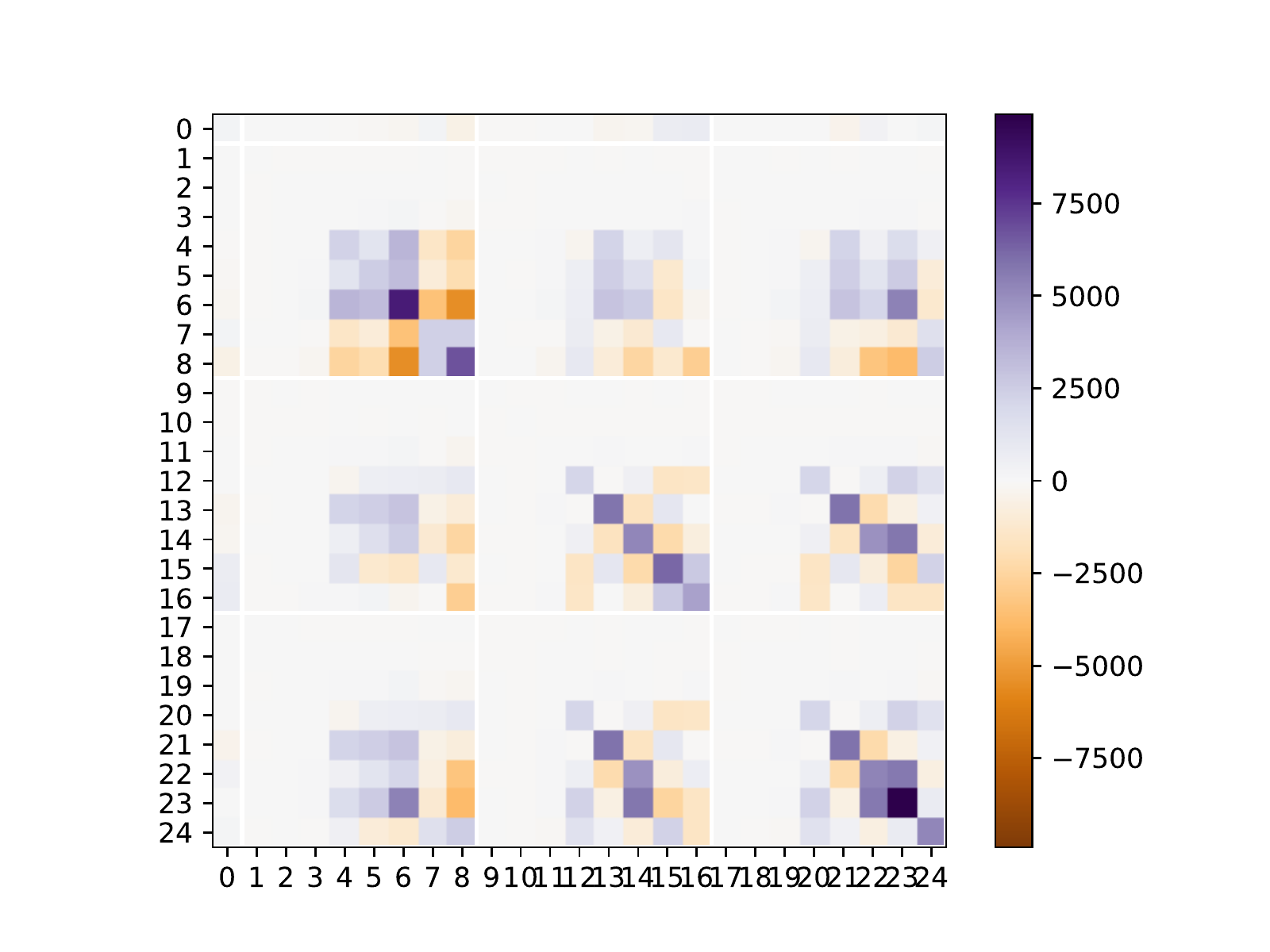}
	\caption{Covariance matrix as heat map.
		Each row and each column correspond to the relative offsets in the scope of the overlayed and aligned set of segments of one cluster.
		White lines are the relative positions of true boundaries in this set.}
	\label{fig:heatmap}
\end{figure}

\subsection{Interpretation of the PCA}\label{sec:interpretation-of-the-pca}

The clusters from the previous step that are suitable for PCA are interpreted in two steps.
First we apply inference rules for field boundaries and then we determine commonly aligned offsets within clusters that likely show additional boundaries.
For any interpretation, we use the common alignment of all segments within a cluster, which we obtain as explained in \autoref{sec:align}.
This results in \emph{relative offsets} that are common throughout one cluster.
Thus, the field-boundary inference rules and the additional conditions for boundaries work on this dissimilarity-aligned segments per cluster.

\subsubsection{Inference Rules}\label{sec:inference-rules}

The covariance shows transitions between related message parts in the byte sequences of a set of segments provided by a cluster.
To get an impression about how the covariance matrix $\covmatrix$ represents such related parts, regard \autoref{fig:heatmap}.
The more intense the color, the stronger the linear dependent variance at this offset.
We use the loadings $\loading$ of \emph{significant} PCs calculated from $\covmatrix$ to test for conditions that represent characteristics of field boundaries.
We deduce two different rules from the typical data type representation in byte streams of network messages.

\paragraph{Rule A}
The first rule is governed by the observation that the data of a field with a common data type exhibits a rise in variance towards the field's end~\cite{kleber_nemesys:_2018}.
\begin{figure}
	\includegraphics[width=\linewidth] 
		{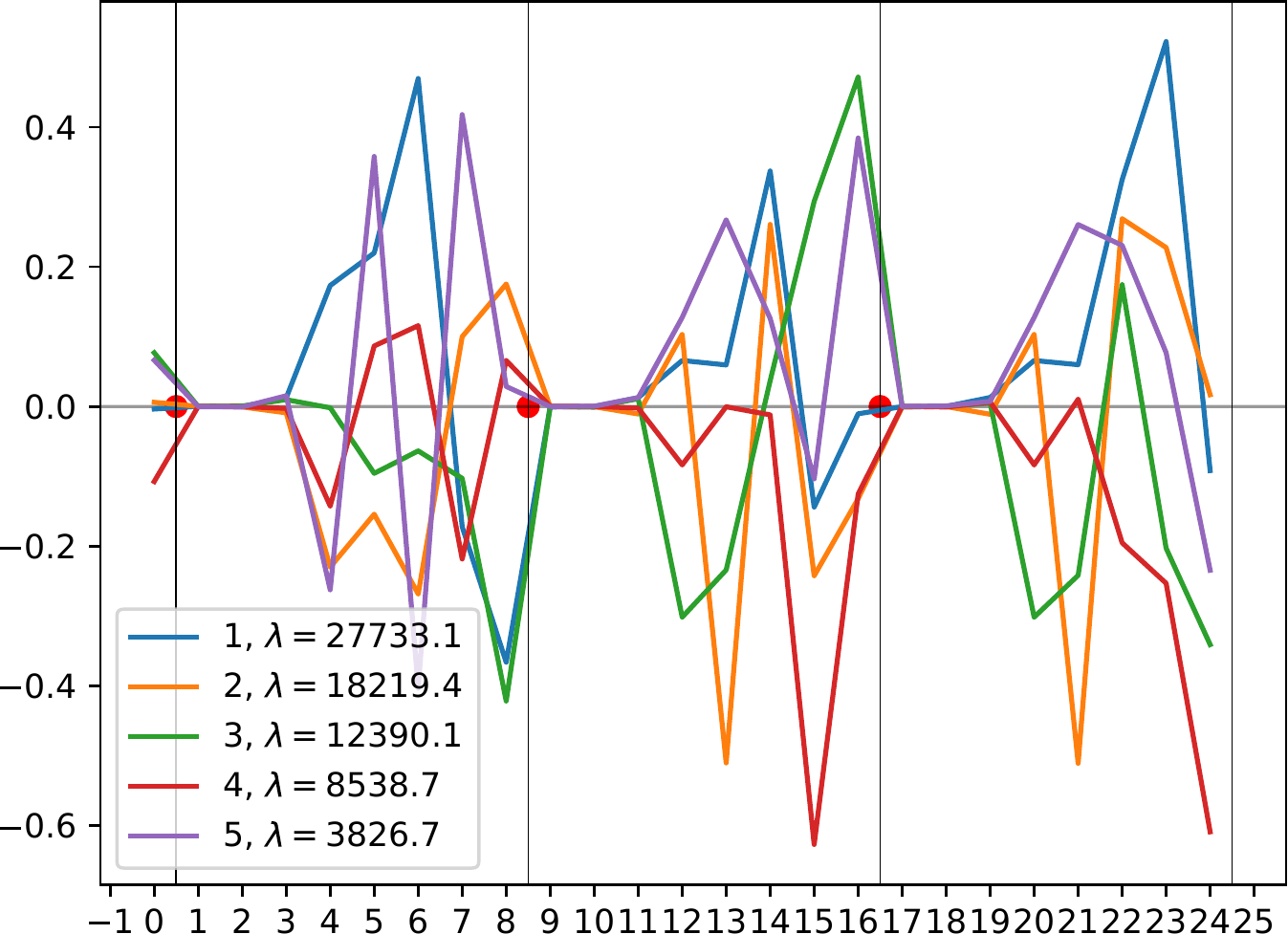}
	\caption{Example for matches of rule A: significant loadings of one cluster (with eigenvalue > 2\,773).
		The dot denotes the matches, the vertical lines the true boundaries.}
	\label{fig:rulea}
\end{figure}
%
Using the same cluster as in \autoref{fig:heatmap} as basis, \autoref{fig:rulea} illustrates this by the loadings of the significant PCs for segments of 25 bytes length.
A PC $i$ is significant if it has an eigenvalue $\score_i$ above 
the threshold $\screeThresh$ 
(\autoref{sec:recursive-clustering}). 
We thus define a significant PC $i$ to relevantly contribute to the variance at a byte position $k$, as such with its maximum loading $m_k > \contributionRelevant$, with
\begin{equation}
\maxContribution_k = \max_{i=1}^n\left<\left|\loading_{i,k}\right|\right>
\label{eq:maxContribution}
\end{equation}
where $n$ is the number of significant PCs $i : \score_i > \screeThresh$.

We now search for a considerable relative drop in the absolute covariance after the peak at the end of a suspected field.
Thus, a field boundary is detected at byte position $k$ if the Boolean expression
\begin{equation}
	\maxContribution_{k-1} > \contributionRelevant 
	\Land \maxContribution_k \leq \contributionRelevant
	\Land \frac{\maxContribution_{k-1} - \maxContribution_k}{\maxContribution_{k-1}} > \pcDeltaMin
\end{equation}
is true, regardless of which $\score_i$ is responsible for the peak. 

\paragraph{Rule B}
The second rule detects the start of a new field in situations where a prolonged sequence of byte positions do not significantly vary and the next byte suddenly peaks in its variance.
%
%
This typically happens at the transition between constant fields or such with few possible values, like enumerations, commands, or flags.

The maximum loading $\maxContribution$ is defined as for Rule A in \autoref{eq:maxContribution} and we introduce the thresholds 
$\relaxedNZ$ to denote loadings that are near zero,
$\relaxedNZlength$ the length in bytes that a loading needs to be below $\relaxedNZ$ to trigger the rule,
and $\relaxedNoteableContrib$ the minimum absolute value of a loading to constitute a notable contribution to the variance.
A variance peak after a prolonged sequence of near constancy is thus defined to be at a byte position $k$ for which the following Boolean expression is true:
\begin{equation*}
	\bigwedge_{j=1}^{\relaxedNZlength} \maxContribution_{k-j} < \relaxedNZ
	\Land \maxContribution_k > \relaxedNoteableContrib
	\Land \frac{\maxContribution_{k} - \maxContribution_{k-1}}{\maxContribution_{k}} > \pcDeltaMin
\end{equation*}
%

We apply both rules to all suitable clusters individually.
We move or add boundaries at the same relative offsets for all segments in one cluster for which any of the rules apply.

\subsubsection{Commonly Aligned Boundaries Conditions}
\label{sec:common-aligned-boundaries-conditions}

To overlay segments of a cluster, we dislocate the segments against each other as described in \autoref{sec:align}, so that the sub-sequence of one segment is at the same relative offset that is most similar to the other segment.

If the overlaying shows a prevalent common boundary in the aligned starts and ends of segments
and if that boundary is missing in only a minority of segments of the cluster,
we cut segments within one cluster at all relative offsets of boundaries that are more common than their direct neighbor, which originated from the raw segmentation or which where detected by component analysis.
The reason, we can detect these errors in the raw segmentation, is that the alignment on Canberra dissimilarity reveals segment relations.
\section{Null-Bytes Transitions}
\label{sec:nulls}

The central goal of our work is to improve the accuracy of the field inference of unknown messages.
PCA requires a preexisting segmentation that supplies comparable data vectors.
We observed that null bytes, produced by unset fields, unused most significant bytes of numbers, or null terminated strings, denote probable field boundaries in binary protocols~\cite{kleber_nemesys:_2018}.


Applied as a refinement for \sys, we improve its raw segments by relocating inferred boundaries near the beginning and end of sequences of nulls:
%
%
(1) If the last bytes before the nulls fulfill the character
heuristic~\cite{kleber_message_2020}, we assume it is a null-termination for the string and allocate the nulls to the end of the character segment;
(2) otherwise, we assign the null-bytes to the following segment, assuming they are the unset most significant bytes of a number value.
%


As an alternative, we propose a novel standalone segmenter that is not based on \sys to compare this as a simpler foundation to apply our PCA afterwards.
Similar to the null-bytes refinement for \sys, the segments that our so-called \emph{Null-Bytes Segmenter} yields are the subsequences of the messages which are delimited by null bytes.
The segmentation is only coarse but still adequate
to prepare the fine-grained field inference by PCA.
%

\section{Implementation}
\label{sec:implementation}

We implemented a proof-of-concept in Python\:3.%
\footnote{\url{https://github.com/vs-uulm/nemesys}}
We require numerous thresholds and other parameters throughout our method, for which
\autoref{tab:parameters} provides an overview.
To empirically determine universal values of the fixed parameters and thresholds our approach relies on, we used \textbf{traces} of the binary network protocols DHCP, DNS, NBNS, NTP, and SMB.%
\footnote{
	Dynamic Host Configuration Protocol (RFC 2131), 
	Domain Name System (RFC 1035), 
	NetBIOS Name Service (RFC 1002), 
	Network Time Protocol (RFC 958), and
	Server Message Block 
}
These traces are publicly available.%
\footnote{%
	DHCP, NBNS, NTP, and SMB extracted from \url{http://download.netresec.com/pcap/smia-2011/};
	DNS extracted from \url{https://ictf.cs.ucsb.edu/archive/2010/dumps/ictf2010pcap.tar.gz}}%
%
%
%
%
%

\begin{table}[b]
	\caption{Parameters and empirically determined values for our algorithms.}
    \label{tab:parameters}
    \footnotesize
	\begin{tabularx}{\linewidth}{@{}Xll@{}}

        \rowcolor[HTML]{CCCCCC}\rule{0ex}{2.5ex}%
		\textbf{Parameter} & \textbf{Task} & \textbf{Value} \\  
		
		\rule{0ex}{2.5ex}%
		Scree threshold & Sub-cluster & $\screeThresh$ \\
		Scree minimum & Sub-cluster & $\screeMinThresh = 10$ \\
		Maximum significant principals & Sub-cluster & $\maxAbsolutePrincipals = 4$ \\
		Significant principals ratio & Sub-cluster & $\principalCountThresh = 0.5$ \\
		Length difference ratio & Sub-cluster & $\maxLengthDeltaRatio = 0.5$ \\
		Minimum cluster size & Sub-cluster & $\minSegLen = 6$ \\
		
		Significant $\loading$-contribution threshold & Field inference A & $\contributionRelevant = 0.1$ \\
		Signific. loading-difference threshold & Field inference A & $\pcDeltaMin = 0.98$ \\
		
		Near-zero threshold & Field inference B & $\relaxedNZ = 0.05$ \\
		Near-zero minimum length & Field inference B & $\relaxedNZlength = 4$ \\
		Notable $\loading$-contribution threshold & Field inference B & $\relaxedNoteableContrib = 0.005$ \\ 

	\end{tabularx}
\end{table}


Both, 
our novel method based on PCA for refining the \sys segmenter and 
our standalone Null-Bytes Segmenter, process raw segments in a chain of refinements.
This concept is similar to the original segmenter's and the refined version of \tyl.
\autoref{tab:refinements} provides an overview of the applied refinements from literature and our own approaches.

The refinement methods introduced by previous work are the simple char detection heuristic in \sys, which we call \emph{MergeCharsv1}, and the advanced char detection heuristic in \tyl, \emph{MergeCharsv2}.
\tyl counts the most frequent segment values and crops these from larger segments.
We call this refinement \emph{CropDistinct}.
\tyl also introduced the splitting of the first segment of each message into fixed chunks, which we denote \emph{SplitFixedv1}.

The primary contributions of this paper are two novel methods for the refinement chain.
The first one are the Null-Bytes Refinement and Segmenter \autoref{sec:nulls}, which we abbreviate by \emph{NullBytes}, and the second one is the application of Principal Component Analysis to guide segment refinements, \emph{PCA} (see \autoref{sec:PCA}).
In preparation of PCA, we propose \emph{EntropyMerge} for merging of \sys segments if they have similar local entropy.
We also add slight improvements for the splitting of the first segment of each message, \emph{SplitFixedv2},
and the handling of cropping char segments, \emph{CropChar}, which is based on \tyl's \emph{MergeCharsv2}.

%
%
%
%

\todo{EntropyMerge, CropChar, and SplitFixedv2 commented out}

\begin{table}[b]
	\caption{Refinement overview for the compared approaches. Our contributions are printed in bold font.}
	\small\centering
	\begin{tabular}{@{}c c c c@{}}
        \rowcolor[HTML]{CCCCCC}\rule{0ex}{2.5ex}%
		\sys & \tyl & \textbf{\pca} & \textbf{NullPCA} \\
        \rowcolor[HTML]{CCCCCC} 
		\emph{Original}~\cite{kleber_nemesys:_2018} & \emph{Baseline}~\cite{kleber_message_2020} & \cite{kleber_nemesys:_2018} + Sec. \ref{sec:BSVA} & Sec. \ref{sec:nulls} + \ref{sec:BSVA} \\
        
        \rule{0ex}{2.5ex}%
		\sys               & \sys                  & \sys                  & --                    \\ 
		MergeCharsv1       & MergeCharsv2          & \textbf{EntropyMerge} & \textbf{NullBytes}    \\ 
		                   & CropDistinct          & \textbf{NullBytes}    & CropChars             \\ 
		                   & SplitFixedv1          & CropChars             & \textbf{PCA}          \\ 
		                   &                       & \textbf{PCA}          & CropDistinct          \\ 
		                   &                       & CropDistinct          & \textbf{SplitFixedv2} \\
		                   &                       & \textbf{SplitFixedv2} &                       \\
	\end{tabular}
	\label{tab:refinements}
\end{table}

Using these refinement methods as elements for a processing chain, we apply our improvements in two different ways that we call \pca and NullPCA.
We mark the placement of our contributions in the novel processing chains by a bold font in \autoref{tab:refinements}.

\section{Evaluation}
\label{sec:evaluation}

Using our proof-of-concept implementation presented in \autoref{sec:implementation}, we evaluate our approach.
We evaluate the quality of the inferred segment boundaries with seven representative binary network protocols.
Beyond this directly measurable effect on the segmentation, we show the impact of the refined segments that our method yields by applying analyses from previous work using the improved segments as starting point.
Thus, we perform message type identification as previous work proposed in conjunction with segments from \sys~\cite{kleber_message_2020}.
Furthermore, we classify field data types using an existing method, which we also proposed to work with segments from \sys~\cite{kleber_network_2022}.

Additionally to the protocols DHCP, DNS, NBNS, NTP, and SMB that we used to select the parameters for our algorithm (\autoref{sec:implementation}), we also use our own traces of the two proprietary protocols Apple Wireless Direct Link (AWDL) and Auto Unlock (AU).
AWDL is a Wi-Fi-based link-layer protocol for peer-to-peer communication.
AU implements a proprietary distance bounding protocol.%
\footnote{\url{https://support.apple.com/en-us/HT206995}}
Both protocols were undocumented until they recently have been manually reverse engineered.
The reverse-engineered specification of AWDL, including a dissector, is publicly available~\cite{stute_one_2018}, and we had access to a private Wireshark dissector of the AU protocol.
Thus, both protocols are typical PRE use cases while we have ground truth available.
As the source of the ground truth, we parse the Wireshark dissectors' output for each message.
All evaluated protocols are binary, while DNS, DHCP, SMB, and AWDL also contain embedded char sequences.
The binary fields of DNS, NBNS, and NTP have fixed length, while DHCP, SMB, AWDL, and AU use a mix of fixed and variable-length fields.
DHCP, DNS, NBNS, SMB, AWDL, and AU support varying numbers of fields in different messages, while NTP has a fixed structure.
Thus, our set of traces represents a wide variety of protocol properties.

We compare the results of the three approaches \tyl-refined baseline, \pca, and NullPCA which we described in \autoref{sec:implementation}.
The \tylrefine is the refinement as described in \tyl and the other two are applications 
with 
and without \sys (\autoref{sec:nulls}), respectively, in conjunction with our novel analysis method using PCA, introduced in \autoref{sec:BSVA}.

\subsection{Message Segmentation}
\label{sec:segmentation}
The immediate effect we expect of our refinement is that the segment boundaries will more accurately match the field boundaries of the protocol specification.
To measure this, we use the Format Match Score (FMS) that we proposed together with \sys~\cite{kleber_nemesys:_2018}.
We apply our refinements to each of our test protocols, calculate the FMS, and discuss the results in this section.
%
%
%
%
%
%
%
The FMS is a measure of correctness of the inference of a message format.
By the FMS, we can quantitatively compare the quality of different inference methods.
The FMS is calculated for each message individually
and we therefore calculate the median of all FMS' for one trace.


\begin{table}[b]
	\caption{Comparison of message segmentation quality using the median values of FMS per protocol trace.}
	\label{tab:fms}
    
    \raggedright
    \footnotesize 
    \textbf{Note:} Numbers printed in bold in the protocol rows are the worst cases discussed in the text and the bold median values at the bottom are the mainly discussed comparison.
	
    \medskip
	\small\centering
	\begin{tabular}{rr|ll|ll|r}
		\multicolumn{2}{l}{\textbf{FMS median}}                              & \multicolumn{2}{|l}{\tylrefine} & \multicolumn{2}{|l|}{\cellcolor[HTML]{777777}\color{white}\pca} & NullPCA
		 \\
		\rowcolor[HTML]{CCCCCC} 
		\textbf{trace}                    & \textbf{msg.s}                   & $\sigma$ \textbf{0.9}      & $\sigma$ \textbf{1.2}     & $\sigma$ \textbf{0.9}     & $\sigma$ \textbf{1.2}      &     \\
		DHCP & 100  & \textbf{0.29} & 0.23 & 0.38 & \textbf{0.37} & \textbf{0.21} \\
		DHCP & 1\,000 & \textbf{0.30} & 0.28 & 0.38 & \textbf{0.35} & \textbf{0.29} \\
		DNS  & 100  & 0.56          & 0.48 & 0.74 & 0.74          & 0.74          \\
		DNS  & 1\,000 & 0.51          & 0.49 & 0.90 & 0.90          & 0.87          \\
		NBNS & 100  & 0.52          & 0.29 & 0.66 & 0.66          & 0.57          \\
		NBNS & 1\,000 & 0.55          & 0.32 & 0.68 & 0.68          & 0.56          \\
		NTP  & 100  & 0.58          & 0.72 & 0.66 & 0.71          & 0.53          \\
		NTP  & 1\,000 & 0.59          & 0.71 & 0.59 & 0.71          & 0.52          \\
		SMB  & 100  & \textbf{0.36} & 0.29 & 0.59 & 0.61          & 0.47          \\
		SMB  & 1\,000 & \textbf{0.39} & 0.30 & 0.56 & 0.52          & \textbf{0.36} \\
		AWDL & 100  & 0.45          & 0.35 & 0.62 & 0.60          & \textbf{0.35} \\
		AWDL & 768  & 0.40          & 0.33 & 0.54 & 0.52          & 0.46          \\
		AU   & 123  & 0.60          & 0.56 & 0.41 & \textbf{0.38} & \textbf{0.23} \\
		\rowcolor[HTML]{CCCCCC} 
		\multicolumn{2}{l}{\cellcolor[HTML]{CCCCCC}\textbf{median}} & \textbf{0.51}     & 0.33             & 0.59             & \textbf{0.61}     & 0.47 \\
	\end{tabular}
\end{table}



Summarized in \autoref{tab:fms}, 
the \tyl~\cite{kleber_message_2020} \tylrefine performs worst for DHCP and SMB, while our PCA refinement, in contrast, stays lower for DHCP and AU, however at a reasonable absolute value of above 0.35 and still improving on \tyl's DHCP results of below of 0.30.
For all other protocols, \pca clearly outperforms the \tylrefine except for NTP where the FMS values are closeby.
The results also show that traces of 1\,000 and 100 messages are almost identical, confirming that decent results can be produced with even small traces.
Compared to the \tylrefine, our \pca yields better results for 100 than for 1\,000 DHCP, SMB, and AWDL messages, showing that it is effective in extracting more structural information even from small traces than the previous approaches.

\sys requires to select a value for $\sigma$ that is dependent of the field lengths expected for a protocol.
The results for the \tylrefine show that the quality is significantly higher on average for $\sigma = 0.9$.
Our goal is to become independent from this parameter, since it is difficult to determine it correctly.
Performing our \pca refinements on \sys segments with different $\sigma$ values, we observe that the results are almost identical for both $\sigma$ values.
The only protocol that declines in quality is AU while all other protocols increase their field correctness in terms of the FMS. 
Assuming that we know nothing about the protocol that helps to select the optimal $\sigma$ value and thus blindly selecting $\sigma = 1.2$ for the \tylrefine, we improve the results by almost 100\,\% using \pca due to its robustness against $\sigma$ changes.

Finally, we used our novel Null-Bytes Segmenter that works without \sys and applied the PCA refinement to it as described in \autoref{sec:implementation}.
The resulting FMS values on average are considerably lower than those of \pca, but similar to the \tylrefine.
This shows the advantage of using \sys as a heuristic method and that the complex effort of refining its segmentation is worthwhile.

%

\subsection{Message Type Identification}
\label{sec:messagetypes}

\begin{table}[b]
	\caption{Message type identification quality measured by precision (P) and recall (R).}
	\label{tab:mti}
	
	\small\centering
	\begin{tabular}{@{}r@{\hspace{8pt}}r||rr|rr||rr@{}}
		&
		&
		\multicolumn{2}{c|}{\tylrefine} &
		\multicolumn{2}{c||}{\cellcolor[HTML]{777777}\color{white}\pca} &
		\multicolumn{2}{c}{NullPCA} \\
		\rowcolor[HTML]{CCCCCC} 
		\textbf{trace} & 
        \textbf{msg.s} & 
		\multicolumn{1}{|c}{\textbf{P}} &
		\textbf{R} &
		\textbf{P} &
		\textbf{R} &
		\textbf{P} &
		\textbf{R} \\
        DHCP & 100  & 0.94 & 0.10 & 0.98 & 0.11 & 0.96 & 0.15 \\
        DHCP & 1\,000 & 1.00 & 0.58 & 0.99 & 0.77 & 1.00 & 0.49 \\
        DNS  & 100  & 1.00 & 0.45 & 0.88 & 0.41 & 1.00 & 0.59 \\
        DNS  & 1\,000 & 1.00 & 1.00 & 1.00 & 0.24 & 1.00 & 0.54 \\
        NBNS & 100  & 0.99 & 0.68 & 0.98 & 0.88 & 0.99 & 0.47 \\
        NBNS & 1\,000 & 0.78 & 0.89 & 0.80 & 0.83 & 0.89 & 0.35 \\
        NTP  & 100  & 0.20 & 0.87 & 0.77 & 0.18 & 0.91 & 0.14 \\
        NTP  & 1\,000 & 0.52 & 0.69 & 0.51 & 0.85 & 0.52 & 0.94 \\
        SMB  & 100  & 1.00 & 0.28 & 1.00 & 0.28 & 1.00 & 0.25 \\
        SMB  & 1\,000 & 0.89 & 0.75 & 0.92 & 0.71 & 1.00 & 0.75 \\
        AWDL & 100  & 1.00 & 1.00 & 1.00 & 0.64 & 1.00 & 0.86 \\
        AWDL & 768  & 1.00 & 0.52 & 1.00 & 0.44 & 1.00 & 0.51 \\
        AU   & 123  & 1.00 & 0.26 & 1.00 & 0.18 & 1.00 & 0.16 \\
		\rowcolor[HTML]{CCCCCC} 
		\multicolumn{2}{l}{\textbf{median}} &
		1.00 &
		\textbf{0.68} &
		0.98 &
		\textbf{0.59} &
		\textbf{1.00} &
		\textbf{0.51} \\
	\end{tabular}
\end{table}

We apply the \tylrefine segmentation to identify message types and compare these results with applying the \tyl message type identification on top of our segment refinements.
\autoref{tab:mti} shows the results measured in classification precision and recall.
They where calculated exactly as described in the \tyl paper~\cite{kleber_message_2020}.
For simplicity, we only use $\sigma = 1.2$ for the \sys-based analyses: the \tylrefine and \pca.
While the precision is almost optimal for all of the compared approaches, the previous and our novel ones,
Our \pca can reach or outperform the \tylrefine only for few traces.

NullPCA has a precision that is on average similar to the others but with a recall that is notably lower.
However, NullPCA is able to outperform the precision for single protocol traces of NTP and SMB that are particularly difficult for both, the \tylrefine and \pca.
This is due to the structure of these protocols that contain sequences of null bytes separating segments which denote the message type.

Our method cannot provide groundbreaking improvements for message type identification.
This shows that a more accurate field approximation is not necessarily improving the message type identification.
NullPCA is an alternative only if the protocol for the most part contains segments that are clearly separated by null bytes.

\subsection{Field Data Type Clustering}
\label{sec:fieldclassification}

\begin{table*}[!]
    \vspace{0.08in}
	\caption{Field data type clustering quality in precision (P), recall (R), noise, and number of unknown segments (unk.).}
	\label{tab:ftc}
	
	\small\centering
	\begin{tabular}{rr|rrrr||rrrr||rrrr}
		& &
		\multicolumn{4}{c}{\tylrefine} &
		\multicolumn{4}{c}{\cellcolor[HTML]{777777}\color{white}\pca} &
		\multicolumn{4}{c}{NullPCA} \\
		\rowcolor[HTML]{CCCCCC} 
		\textbf{trace}                    & \textbf{msg.s} &
		\textbf{P} &
		\textbf{R} &
		\textbf{noise} &
		\textbf{unk.} &
		\textbf{P} &
		\textbf{R} &
		\textbf{noise} &
		\textbf{unk.} &
		\textbf{P} &
		\textbf{R} &
		\textbf{noise} &
		\textbf{unk.} \\
        DHCP & 100  & 0.86 & 0.51 & 87   & 187  & 0.71 & 0.24 & 125 & 79   & 0.83   & 0.30   & 121   & 75    \\
        DHCP & 1\,000 & 0.86 & 0.19 & 275  & 972  & 0.95 & 0.47 & 166 & 192  & 0.87   & 0.78   & 29    & 93    \\
        DNS  & 100  & 0.88 & 0.21 & 10   & 23   & 0.97 & 0.50 & 27  & 4    & 1.00   & 0.53   & 31    & 5     \\
        DNS  & 1\,000 & 0.87 & 0.21 & 31   & 30   & 0.99 & 0.97 & 20  & 5    & 1.00   & 0.92   & 50    & 6     \\
        NBNS & 100  & 0.99 & 0.85 & 1    & 19   & 0.99 & 0.52 & 13  & 19   & 0.99   & 0.89   & 5     & 4     \\
        NBNS & 1\,000 & 0.99 & 0.80 & 12   & 34   & 0.99 & 0.58 & 4   & 14   & 1.00   & 0.94   & 5     & 290   \\
        NTP  & 100  & 0.81 & 0.75 & 93   & 25   & 0.99 & 0.15 & 108 & 105  & \multicolumn{4}{c}{knee detection failed} \\
        NTP  & 1\,000 & 0.69 & 0.30 & 2\,128 & 189  & 0.89 & 0.21 & 731 & 1\,105 & 0.44   & 0.84   & 36    & 1\,105  \\
        SMB  & 100  & 0.84 & 0.12 & 128  & 181  & 0.77 & 0.22 & 85  & 80   & 0.80   & 0.20   & 98    & 72    \\
        SMB  & 1\,000 & 0.37 & 0.02 & 692  & 1\,549 & 0.67 & 0.34 & 268 & 664  & 0.89   & 0.20   & 439   & 551   \\
        AWDL & 100  & 0.74 & 0.05 & 270  & 393  & 0.63 & 0.10 & 192 & 206  & 0.73   & 0.17   & 22    & 193   \\
        AWDL & 768  & 0.80 & 0.16 & 261  & 2\,205 & 0.45 & 0.01 & 929 & 703  & 0.84   & 0.50   & 44    & 1\,235  \\
        AU   & 123  & 1.00 & 0.06 & 662  & 352  & 0.99 & 0.57 & 193 & 121  & 0.99   & 0.45   & 89    & 880   \\
        \rowcolor[HTML]{CCCCCC} 
        \textbf{median} &
          \textbf{} &
          0.86 &
          0.26 &
          90 &
          108 &
          \textbf{0.96} &
          \textbf{0.41} &
          97 &
          \textbf{80} &
          \textbf{0.89} &
          \textbf{0.78} &
          36 &
          \textbf{75}
	\end{tabular}
\end{table*}

The second analysis based on the refined segments that we use as evaluation aspect is the clustering of field data types from \sys' message segments~\cite{kleber_network_2022}.
\autoref{tab:ftc} shows our evaluation results by measuring precision, recall, the number of segments that are considered noise by the clustering algorithm, and the number of unknown segments.
Segments are considered unknown if no match to any field from the ground truth is possible that would allow to determine the field type for the segment.
Thus, this value denotes one aspect of the deviation of the heuristic segmentation from the true field structure of the protocol.

Comparing the \tylrefine to \pca, the precision stays similar for most protocols, but we can increase the recall regarding the median of all protocols by 57\,\%.
The number of unknown segments are reduced considerably as expected due to improved segment accuracy.
As an exception, the increased unknowns of NTP account for the low recall of the traces.
The larger AWDL trace yields a high amount of noise and thus the lowest precision and recall of all traces.
The protocol is highly complex and eludes our heuristic.
However, on average, we can improve precision and recall while reducing unknown segments significantly.

Our other proposal, NullPCA, shows tripled recall and reduced noise and unknown segments compared to the \tylrefine.
This comes at the cost of a slightly decreased precision, however.
Also, the usage of null bytes as primary segment detection mechanism leads to the loss of details in the boundaries of tightly packed fields that \sys can discern.

%


\section{Conclusion}
\label{sec:conclusion}

In this paper, we presented a chain of refinements for two segmenters, \sys and our simple null-byte transition detection.
Besides the static rule-based algorithms 
\emph{EntropyMerge}, 
\emph{NullBytes},
\emph{CropChars},
\emph{CropDistinct}, and
\emph{SplitFixed}, 
we introduce the novel, dynamic \emph{PCA} method.
It measures the byte-wise variance of segment contents to increase the accuracy of the field boundary detection in unknown binary protocols over previous work.
Opposed to its common application in classification tasks, we utilize PCA directly to determine linearly dependent variance and use a novel interpretation of these results to detect field boundaries from a rough preliminary segmentation.

Our evaluation shows improved field inference quality of up to 100\,\% for most protocols.
Moreover, it renders any parameter selection by guessing unnecessary.
PCA is effective in extracting more structural information even from small traces than the previous approaches.
Thus, our method notably improves the field accuracy as measured by FMS over the previous segmenters and their refinements.

As we discuss within our evaluation, our simple Null-Bytes Segmenter, combined with PCA, provides slightly less accuracy than refined \sys segments.
Despite the original assumption of PCA that it can correct off-by-one errors in particular, the PCA refinement still works well enough for segments generated only from the slicing of null-bytes.
However, while more complex, using \sys segments as basis for further analyses outperforms this simpler segmenter.
Thus, we conclude that the improved inference accuracy makes the relatively complex refinement of segments from \sys worthwhile.

Encrypted messages cannot be inferred by our method directly and require obtaining of plain-text traces.
Furthermore, PCA can only detect linearly dependent correlations and fails for random and non-linearly correlated message values.
A larger set of input data improves the accuracy only slightly, while the exponential memory complexity of our refinement limits the practical applicability to traces of a little over one thousand messages.
Thus, small traces that contain high variability of message values are preferable for our method.
We propose for future work to find more sophisticated rule sets for deducing boundaries from relations between PCs.
This will improve the accuracy of the field boundary detection.

The inference of the message format is a crucial task in protocol reverse engineering.
Often performed during security assessments, it provides knowledge about the placement of field boundaries in protocol messages,
which is necessary to correlate information to field values.
Among other use cases, it helps to determine portions of the message to inspect, e.\,g., by Fuzz testing,
to identify features for the fingerprinting of protocols and to train intrusion detection systems.
Recent security analyses~\cite{stute_one_2018,kroll_aristoteles_2021} have shown the value that this kind of method adds to the arsenal of a security analyst.

\section*{Acknowledgment}
We would like to thank Milan Stute for his support and the AWDL traces, as well as Steffen Klee for providing us with a Wireshark dissector and traces for Apple's Auto Unlock (AU) protocol.

\printbibliography


\end{document}
